# Near-band-gap photo-induced nuclear spin dynamics in semi-insulating GaAs: Hyperfine- and quadrupolar-driven relaxation

*Yunpu Li[1,*], Jonathan P. King[2,†], Jeffrey A. Reimer[2], and Carlos A. Meriles[1]*

[1]Department of Physics, CUNY-City College of New York, New York, NY 10031, USA

[2]Department of Chemical and Biomolecular Engineering, University of California at Berkeley, Berkeley, CA 94720, USA.

**ABSTRACT**

Understanding and manipulating spin polarization and transport in the vicinity of semiconductor-hosted defects is a problem of present technological and fundamental importance. Here, we use high-field magnetic resonance to monitor the relaxation dynamics of spin-3/2 nuclei in semi-insulating GaAs. Our experiments benefit from the conditions created in the limit of low illumination intensities, where intermittent occupation of the defect site by photo-excited electrons leads to electric field gradient fluctuations and concomitant spin relaxation of the neighboring quadrupolar nuclei. We find indication of a heterogeneous distribution of polarization, governed by different classes of defects activated by either weak or strong laser excitation. Upon application of a train of light pulses of variable repetition rate and on/off ratio, we uncover an intriguing regime of mesoscale nuclear spin diffusion restricted by long-range, non-uniform electric field gradients. Given the slow time scale governing nuclear spin evolution, such optically-induced polarization patterns could be exploited as a contrast mechanism to expose dark lattice defects or localized charges with nanoscale resolution.

[*] *Present address*: Philips Research North America, 345 Scarborough Road, Briarcliff Manor, NY 10510, USA.
[†] *Present address:* Department of Chemistry, University of California at Berkeley, Berkeley, CA 94720, USA.

## I. INTRODUCTION

Multi-spin clusters comprising single or multiple electrons trapped in a solid-state matrix and neighboring nuclear spins provide a fascinating model system in mesoscale physics. At the interface between the quantum and the classical, these "spin complexes" are of interest both from fundamental and applied standpoints as they may serve as the platform for scalable quantum information processing in the solid state[1]. Examples of nuclear-electron spin complexes include those formed within self-assembled semiconductor quantum dots[2], in the vicinity of crystal defects such as the Nitrogen-Vacancy center in diamond[3], the di-vacancy and Silicon-Vacancy centers in SiC[4,5], substitutional phosphorous[6] and bismuth[7] in silicon, rare-earth ions[8], and others[9].

Systems formed by trapped photo-excited electrons and neighboring nuclear spins in semiconductors form a sub-class of spin complexes with interesting properties. For example, because these defects often change from paramagnetic to magnetically-inactive once the trapped photo-electron recombines, nuclear spins in the vicinity are inherently protected against electron-spin-induced decoherence. On the other hand, since the spin orientation of photo-excited electrons can be often controlled through light helicity, electron recombination at the defect site also provides an efficient mechanism of electron-nuclear spin polarization transfer. Indeed, "optical pumping" of spin orientation in semiconductors is a well-known technique that is already exploited to generate highly polarized nuclear spin ensembles in bulk crystals[10,11]. At the nanoscale, these same principles have been adapted to the nuclear bath within semiconductor quantum dots, where spin order has been investigated, e.g., as a route to mitigate nuclear-induced decoherence of the electron spin[12,13], control a two-electron qubit[14], tune the polarization of quantum-dot-emitted photons[15], or extract information on local strain[16].

Here we use light-assisted, high-field nuclear magnetic resonance (NMR) to probe the dynamics of nuclear spin relaxation and transport near crystal defects in semi-insulating (SI) GaAs. Our experiments capitalize on the unique conditions created at low illumination intensities, where competing mechanisms of nuclear spin relaxation coexist. Upon slight variation of the laser power we find major changes in the corresponding NMR spectra that we attribute to contributions from nuclear spins near two different classes of defects. By monitoring the NMR response to controlled cycles of light-induced relaxation and nuclear spin diffusion, we find an unanticipated regime of nuclear spin dynamics in which polarization transport is strongly restricted due to non-uniform electric field gradients (EFG) thereby forming a spin transfer blockade. These observations suggest the formation of highly heterogeneous distributions of nuclear spin polarization at the nanoscale, which, if confirmed, could be used to spatially map out localized charge or non-fluorescing defects within the semiconductor host.

## II. EXPERIMENT

We address $^{69}$Ga and $^{71}$Ga nuclear spins in SI GaAs using a custom-built NMR probe immersed within a wide-bore, 9.4 T magnet (field homogeneity better than half a ppm) associated with a triple channel NMR spectrometer[17]. Optical excitation is carried out via a tunable laser beam steered into the magnet bore through the optical windows of a flow cryostat in a direction parallel to the magnetic field. A fast mechanical shutter commanded by the spectrometer controls the illumination timing. The GaAs sample — in the form of a thin wafer — is oriented within the probe so that the normal to its surface coincides with the direction of light propagation. The laser beam impinges on the GaAs crystal through an opening in the NMR pick-up solenoid; the total illuminated area is approximately 1 mm$^2$. The sample is supported within the solenoid by a large sapphire mount acting as a heat sink. Apiezon grease at the interface between the sample and the mount is used to guarantee good thermal contact. For future reference, we note that during cooling the grease tends to disproportionately shrink thus resulting in a long-range crystal strain, uniform on the scale of the illuminated area.

The system we investigated is a 350-μm-thick SI-GaAs crystal from American Crystal Technologies. This sample belongs to a lot identical to that used in prior studies[17-21]: It has resistivity greater than $10^7$ Ω·cm, mobility greater than 6000 cm$^2$/(V·s), and the surface orientation is [100]. Secondary ion beam spectrometry (SIMS) data reveal the presence of S impurities at a concentration of $7*10^{14}$; other impurities including H, C, O, Si, Cr, Mn, and Fe are either absent or below the SIMS detection limit. The interrogation protocol is an extension of that used in the past[10]: After canceling thermal polarization via $N_S$ radio-frequency (rf) pulses, we expose the sample to a predefined sequence of light-on/light off intervals (see below); subsequently, we excite nuclear spins with a resonant π/2-pulse and record the ensuing free induction decay (FID). All experiments reported herein are conducted at 6.5 K. To compensate for $T_1$ relaxation throughout the crystal (most of which is never illuminated), we subtract the signal obtained from a similar protocol in which the light beam is suppressed at all times. Given the working temperature range and the high crystal purity, we find that this "in-the-dark" contribution is consistently small (<10 %) such that it can be easily and reproducibly subtracted from the raw data.

## III. RESULTS AND DISCUSSION

### III.A. Hyperfine versus quadrupolar relaxation

Unlike the typical optically-pumped NMR (OPNMR) study — where the laser power is adjusted so as to provide maximum signal — the present experiments are conducted in the regime of low illumination intensity. In this limit, two mechanisms of nuclear spin relaxation co-exist: The first and best known mechanism stems from hyperfine-mediated spin transfer at the recombination site between the optically polarized photo-electron and neighboring nuclei. As long recognized, this channel of nuclear spin relaxation is key to inducing helicity-dependent nuclear spin order, and thus forms the foundation of OPNMR[22]. More recently a second mechanism was proposed[23] and verified experimentally[20,21] whereby fluctuating electric field gradients in the vicinity of the intermittently photoelectron-occupied defects lead to nuclear spin relaxation via a quadrupolar mechanism. Such fluctuations are inherently insensitive to light helicity, such that quadrupolar relaxation always drives the nuclear spin ensemble towards equilibrium at the applied magnetic field and temperature. As shown previously[20,21] the dominating mechanism can be experimentally dictated using, for example, laser light intensity to change the fractional photoelectron occupation of the defect site to a higher value (driving hyperfine-mediated "optical pumping") or to a lower value (driving quadrupolar relaxation to thermal equilibrium).

Besides their distinct response to laser power and helicity, quadrupolar and hyperfine dynamics exhibit a markedly dissimilar dependence on the distance to the defect. For example, in a model where the recombination site is a shallow donor[23], hyperfine relaxation is most effective at (or in the immediate vicinity of) the defect, but decays exponentially over a distance comparable to $a_0$, the trapped electron Bohr radius (approximately 10 nm in GaAs[24]). Conversely, under quadrupolar relaxation the probability of a nuclear flip via electric field gradient fluctuations vanishes at the defect site and then grows with the radial distance[23] to slowly decay over a distance $\sim 3a_0$. While this trend is shared by systems other than shallow donors, the degree of charge localization in the class of defects active in a given crystal is expected to play a major role in defining the dominant nuclear spin relaxation channel. In particular, we note that deep defects are characterized by higher electron localization, implying that they are associated with stronger electric field gradients and correspondingly enhanced quadrupolar relaxation.

Fig. 1b shows a series of $^{69}$Ga and $^{71}$Ga NMR spectra for different illumination intensities upon application of the standard 'optical pumping' (OP) protocol *Sat-$T_L$-$T_D$-π/2-Acq* (Fig. 1a). Here '*Sat*' denotes a train of $N_S$ equally-spaced π/2-pulses to eliminate any pre-existing sample magnetization, and $T_L$ and $T_D$ respectively indicate the duration of the illumination and dark intervals preceding signal acquisition. As we increase the laser power from about 200 μW to 10 mW, we observe a transformation of the NMR signal from a positive singlet to a negative triplet. To interpret this progression we begin by adopting a spin temperature point of view: After the saturation pulse sequence establishes an infinite temperature in the bath of nuclear spins, quadrupolar relaxation drives spins proximate to the defect to a small, positive temperature whereas hyperfine relaxation results in proximate spins having a large negative temperature (for $σ^+$ helicity). Polarization transport via nuclear dipole-dipole couplings then drives bulk spins towards an effective spin temperature that results from the prevailing relaxation mechanism. Within this framework, one is tempted to interpret the observed progression as a gradual transition from quadrupolar- to hyperfine-dominated nuclear spin relaxation: At low illumination intensities, photo-electron-induced fluctuations of the electric field gradient accelerate relaxation towards thermal equilibrium as manifested by the positive sign of the resulting NMR spectrum. As the number of photo-electrons increase, the role of fluctuations in driving relaxation diminishes and an inverted, hyperfine-induced NMR signal is observed, as expected for illumination with σ⁺ helicity[10,22].

The transformation of the NMR line shape poses, however, an intriguing problem: In our experiments we are able to spectroscopically identify contributions from bulk nuclear spins owing to the long-range strain introduced in the sample during cooling (see section II). Because the lattice distortion created by this process is uniform over macroscopic distances, bulk spins induce an NMR spectrum in the form of a triplet (all active isotopes in GaAs are spin-3/2), whose splitting is dictated by the existing long-range strain. In Fig. 1b, this is precisely the pattern emerging at higher illumination intensities, including the asymmetry in the amplitude of the satellites, already exploited to estimate the average nuclear spin temperature throughout the bulk of the crystal[18]. Comparison with the low laser power spectra shows that at 200 μW, the otherwise sharp satellites have virtually vanished, and the NMR signals of both Ga isotopes take the form of single peaks at (approximately) the frequency of the central transition. These features are indicative of polarized nuclear spins in the vicinity of a defect where a spatially varying electric field gradient broadens the $|\pm 3/2\rangle \leftrightarrow |\pm 1/2\rangle$ transitions without affecting the central peak (at least to first order). We note that since the illumination time both at high and low laser powers is the same, the time scale for nuclear spin diffusion in these experiments remains unchanged. Hence, in a model where the dominant relaxation mechanism near a defect gradually evolves from quadrupolar- to hyperfine-driven, the absence of satellite transitions at low laser power cannot be reconciled with the observation of bulk magnetization at higher illumination intensities.

**III.B. Rationalizing OPNMR in SI-GaAs: Two types of defect at work?**

While the exact nature of the crystal sites responsible for OP in SI GaAs is unknown, our description above builds on the implicit idea (common to most OP studies) that a *single* class of defect dominates nuclear spin relaxation. Instead, we propose that two groups of defects are responsible for our observations: One type of defect is a *deep* defect characterized by a long-range (tens of nm), non-uniform electric field gradient. The higher localization associated with deep defects gives rise to strong electric field gradients and thus efficient quadrupolar relaxation. This strong non-uniform gradient indirectly affects hyperfine-driven relaxation owing to the comparatively shorter range of the hyperfine interaction. Therefore, deep defects are efficient sources of quadrupolar relaxation but lead to only minor hyperfine-induced magnetization. By contrast, the second group is formed by defects that we loosely characterize as *shallow* in the sense that a trapped electron is considerably less localized. In this latter group the degree of delocalization is sufficient to diminish the electric field gradient without making the Fermi-contact contribution to the hyperfine coupling exceedingly weak (as it would be for completely free electrons). Therefore, shallow defects cause comparatively weaker quadrupolar relaxation but are efficient sources of hyperfine relaxation.

Within this framework the results of Fig. 1 can be re-interpreted as the transition between two limit regimes where the nuclear spin dynamics is controlled by one class of defect or the other: Photo-electrons originating from weak laser illumination preferentially populate deep defects, hence leading to quadrupolar relaxation and a positive, broad NMR singlet. As the laser power increases, fuller occupancy of the deep defects progressively quenches quadrupolar relaxation and shallow defects become gradually relevant as manifested by the emergence of a negative triplet. The observation that bulk spins are polarized selectively from shallow sites but not from deep sites (at least to within the time scale of the this experiment), suggests that nuclear spin diffusion dynamics are markedly different near the two different defects.

Fig. 2a introduces a modified excitation protocol explicitly conceived to probe the interplay between nuclear spin relaxation and diffusion. After eliminating thermal polarization via $N_S$ radio-frequency (rf) pulses, we parse the illumination interval into a train of $N_L$ light pulses of duration $t_L$ separated by time intervals $t_D$ in the dark. These dark intervals provide opportunity for nuclear spin diffusion in the absence of light-induced quadrupolar or hyperfine relaxation. The $^{69}$Ga and $^{71}$Ga NMR signals obtained upon application of the above procedure are presented in Fig. 2b (left and right panels, respectively). As with the data shown in Fig. 1, the circularly polarized beam ($\sigma^+$) is on for a total illumination time $T_L = N_L \times t_L = 30$ s, yet this illumination time is intercalated with dark intervals to allow for different regimes of relaxation-diffusion. We note that the series can be thought of as a generalization of the typical OP protocol[10] ($N_L=1$) and thus an extension of the experiments in Fig. 1 for a fixed, intermediate illumination intensity.

Though not a necessary condition (see below), the experiments in Fig. 2 are conducted so that the total dark time $T_D = N_L \times t_D = 300$ s remains unchanged. This condition means that the length scale for nuclear spin diffusion, defined by the diffusion range $r_T = \sqrt{(T_D + T_L)D}$, where $D$ denotes the diffusion coefficient, is the same throughout the series. Remarkably, when we increase $N_L$ from a single pulse to greater values, we witness a major signal transformation, ultimately exhibiting an almost complete sign reversal of the amplitude of the central transition while leaving the satellites virtually unchanged (at least until $N_L$ is sufficiently large). The spectra for $N_L=1$ in the series are virtually identical to those obtained after 30 s of illumination without any wait time in the dark (black traces) thus confirming that potential contributions arising from bulk spin-lattice relaxation can be safely cancelled on this time scale ($T \equiv T_D + T_L = 330$ s)

The results depicted in Fig. 2 have interesting consequences. The experimental protocol has been adjusted so that the laser excitation power produces an inverted triplet under continuous optical pumping (Fig. 2b, spectra labeled *Ref* and $N_L$=1). The appearance of positive, satellite-free NMR signal with an increasing number of dark intervals indicates that our protocol selectively influences the dynamics of nuclear spins near deep defects. We surmise that this is the result of two contributing factors: On the one hand, nuclear spins directly influenced by EFG fluctuations likely reach the limit polarization associated with quadrupolar relaxation (~$5\times10^{-4}$ at 9.4 T and 6.5 K) on a time scale much shorter than $T_L$, which means that fast spin diffusion near the defect is required to avoid a polarization transfer "bottleneck". However, exactly the opposite condition is true near deep defects: The nuclear 'flip-flops' underlying spin diffusion are energy-conserving, meaning that polarization transport is possible so long as the energy splittings between neighboring nuclei are matched. For a spin-3/2 system at high magnetic field, quadrupolar interactions shift the $|\pm 3/2\rangle \leftrightarrow |\pm 1/2\rangle$ transition frequencies according to the nucleus quadrupolar moment and the local electric field gradient, but leave the transition frequency $|1/2\rangle \leftrightarrow |-1/2\rangle$ unchanged (to first order). Therefore, the EFG non-uniformity near the deep defect suppresses flip-flops other than those between nuclei in the $|\pm 1/2\rangle$ states, thus rendering spin diffusion effectively slower. We underscore in this context the extreme sensitivity of spin diffusion to energy mismatch where the ~2 kHz homonuclear contribution to the dipolar line width suggests near-complete flip-flop quenching between spins in the states $|+3/2\rangle$ ($|-3/2\rangle$) and $|+1/2\rangle$ ($|-1/2\rangle$) for shifts of the satellite transitions greater than just ~20 ppm (the Larmor frequency is ~100 MHz at 9.4 T).

As shown in Fig. 3, the transition between spin relaxation mechanisms with illumination intensity is also accompanied by a change in the relative impact of $N_L$ on the outcome of the optically-assisted NMR sequence. Comparison of, e.g., Figs. 3b and 3f shows that the large fractional change observed at low laser powers (where quadrupolar relaxation through deep defects is dominant) virtually disappears at ~20 mW (the regime of hyperfine relaxation near shallow defects). Nuclear spin dynamics at both shallow and deep defects is now seen as a complex interplay between the magnitude of nuclear light-induced polarization at the site, spin transport in its vicinity, and spin transport into the bulk. As we transition to hyperfine-dominated nuclear relaxation at shallow sites, unobstructed spin diffusion to the bulk keeps near shallow-defect spins from reaching the almost complete order required to produce a bottleneck. In the absence of mechanisms restricting the polarization transfer from the defect, redistribution of bulk magnetization during a given dark interval cannot influence the NMR signal, and thus the spectrum amplitude or shape remains unchanged (the regime traditionally found in OPNMR).

### III.C. Polarization transport near deep defects

Additional clues on the spin dynamics near deep defects can be found in the results of Fig. 4a through 4d. Here we measure the $^{69}$Ga NMR response as a function of $t_D$ — the dark time interval during a light-on/light-off cycle — for three different values of $t_L$. Throughout this set of observations, the total illumination time $T_L$ remains constant (30 s) and the number $N_L$ of illumination cycles in each curve can be calculated as the ratio $T_L/t_L$. Unlike the experiments of Fig. 2 — summarized for reference in Fig. 4e — this class of observations takes place over a varying duration $T \equiv T_L + T_D = T_L(1 + t_D/t_L)$. First we note that in the limit of long $t_D$ times, spin-diffusion-driven depolarization during the dark interval in a given on/off cycle keeps nuclear spins in the defect vicinity away from the regime of polarization transfer blockading. As indicated before, maintaining near-defect nuclear spins closer to an unpolarized state readies the system for a new illumination cycle, effectively enhancing the impact of quadrupolar relaxation (note in Fig. 4a the relatively large positive NMR signals for $t_D$ long). In particular, we find that for $t_L$=5 ms — corresponding to $N_L$=6x10$^3$ — the signal exhibits a non-linear growth, asymptotically approaching a limiting positive value. We associate this value with the maximum polarization emerging from light-induced quadrupolar relaxation during $T_L$ in the hypothetical regime of infinitely fast spin diffusion, i.e., in the limit where spin transfer blockading is completely averted. Because all curves in Fig. 4a share the same illumination time $T_L$, we surmise that the same limiting signal should be reached in the other two cases for sufficiently long $t_D$ (provided near-defect nuclei remain nearly unpolarized during the light-on intervals $t_L$=10, 50 ms). On the other hand, and given the large signal amplitudes (comparable to those observed at high illumination intensities), we reason that the concentrations of deep and shallow defects are of the same order of magnitude, suggesting a geometric model for the competing relaxation processes at the two defects. We emphasize that the presented data must be considered free from $T_1$ relaxation in the bulk of the crystal, a contribution (of order 5-15 %) already subtracted in each case upon conducting a 'light-off' experiment of the exact same duration (see section II).

Because nuclear spin diffusion is active during both bright and dark intervals, longer $t_D$ times have a direct impact on the volume of polarized spins $V_T = 4\pi r_T^3/3$, where, as before, $r_T = \sqrt{DT} \approx \sqrt{DT_D}$ defines the range influenced by polarization transport from the defect during the experiment time $T$ (approximately equal to $T_D$ when $t_D \gg t_L$). For example, using the spin diffusion constant[25] $D$~10 nm$^2$/s, we find $r_T \approx 50$ nm (200 nm) for $T_D$ = 300 s (4800 s); in the case $t_L$ = 5 ms we reach this range when $t_D$ = 50 ms (800 ms). We caution that these estimates depend on the value assumed for the diffusion constant $D$, which, as observed in prior studies[26], can decrease by more than an order of magnitude near defects. Interestingly, we find that while the signal amplitude at the frequencies of the satellites in the NMR spectra exhibit a constant, negative amplitude when $T_L$ and $T_D$ are both fixed (Fig. 4e), they gradually turn to positive for longer total dark times (Fig. 4d). This observation suggests that spin diffusion transports nuclear polarization from regions near the defect — where the EFG is heterogeneous at the nanoscale — to the bulk of the crystal — where the gradient turns to uniform on longer length scales. However, even for the longest experimental times ($T \sim 5 \times 10^3$ s for $t_L$=5 ms and $t_D$=0.8 s), the satellites in the resulting NMR spectrum are not well-defined (Fig. 4b) indicating that bulk nuclear spins have not yet been completely reached.

## IV. CONCLUSIONS

Our observations shed new light on the dynamics of nuclear spin polarization by photo-excited electrons in GaAs, and arguably in all semiconductor systems where one or more nuclear isotopes have spin number greater than 1/2. Crucial to our experiments is the regime of low illumination intensities, where quadrupolar-driven nuclear spin relaxation gives rise to unexpected, intriguing phenomenology. Owing to the spectral differences in the NMR response of near-defect and bulk nuclear spins, we identify two classes of recombination centers influencing nuclear spin relaxation. One class comprises shallow defects preferentially inducing hyperfine-driven nuclear spin polarization, whereas the other group is formed by deep centers mostly responsible for quadrupolar relaxation. In this latter case, we find that bottleneck effects associated with restricted spin diffusion and the limited polarization produced by thermal relaxation around these defects considerably influences the resulting spin dynamics. By fractioning the illumination time into low-duty cycles of light-on/light-off intervals we track the evolution of the spin polarization over long, mesoscale distances and find that the defect-centered EFG has an effective range exceeding several tens of nanometers. Because such a distance is an order of magnitude greater than that expected solely from strain in isolated defects, we surmise that the deep defects are charged[27] and/or have the form of extended clusters (e.g., arsenic aggregates extending over 50-200 nm are known to be common in SI GaAs[28]).

The above findings force us to re-examine the notion of 'bulk' polarization in the limit of hyperfine-dominated relaxation, the standard regime in OPNMR[10]. While under strong illumination deep defects do not contribute substantially to the overall nuclear magnetization, its presence is still expected to influence the way hyperfine-driven polarization diffuses throughout the crystal lattice. The result is a heterogeneous distribution of nuclear magnetization characterized by unpolarized pockets with center at deep defects (Fig. 4f). Worth highlighting here is the relatively long range of the mechanisms at play, and thus the inadequacy of optical detection protocols, limited by hyperfine interactions, to probe considerably smaller volumes. Experiments combining, for example, optical pumping and magnetic resonance force microscopy[29] (MRFM) would be ideal to map out this mesoscale texture and more directly expose the differences between the emergence of nuclear polarization near one class of defect or the other. By the same token, such experiments could shed light on the exact nature of these defects, which, remarkably, still remains unknown. In this regard, we note that our observations parallel in many ways the interplay between deep and shallow defects responsible for the low electrical conductivity of SI GaAs[30]. Further studies will be necessary, however, to assign the deep and shallow sites to a particular type of defect.

Our experiments can be extended in interesting ways. For example, one could articulate the present cycles of optically-induced relaxation and spin diffusion with radio-frequency control pulses timed to shape the resulting nuclear polarization as it spreads away from the defect. In particular, one can imagine creating 'onion-like' patterns in which the polarization successively changes sign or amplitude over concentric layers of predefined thickness. The formation of such mesoscale polarization structures could be monitored via the NMR line shifts created by the concomitant long-range fields[31] or could be imaged directly using MRFM[29]. Because the "pitch" in these structures is ultimately influenced by the nuclear spin relaxation mechanism, such experiments could serve as a platform for significant applications, including the detection of trapped charges or the spatial characterization of hyperfine interactions.

## ACKNOWLEDGEMENTS

We thank Melanie Drake for her assistance in obtaining the SIMS data. We also acknowledge support from the National Science Foundation through grant NSF-1106288.

## REFERENCES

[1] *Quantum Information Science and Technology Roadmap,* 2004.

[2] See, for example, M.N. Makhonin, K.V. Kavokin, P. Senellart, A. Lemaître, A.J. Ramsay, M.S. Skolnick, A.I. Tartakovskii, "Fast control of nuclear spin polarization in an optically pumped single quantum dot", *Nature Mat.* **10**, 844 (2011).

[3] L. Childress, M.V. Gurudev Dutt, J.M. Taylor, A.S. Zibrov, F. Jelezko, J. Wrachtrup, P.R. Hemmer, M. D. Lukin, "Coherent Dynamics of Coupled Electron and Nuclear Spin Qubits in Diamond", *Science* **314**, 281 (2006).

[4] W.F. Koehl, B.B. Buckley, F.J. Heremans, G. Calusine, D.D. Awschalom, "Room temperature coherent control of defect spin qubits in silicon carbide", *Nature* **479**, 84 (2011).

[5] D. Riedel, F. Fuchs, H. Kraus, S. Vath, A. Sperlich, V. Dyakonov, A.A. Soltamova, P.G. Baranov, V.A. Ilyin, G.V. Astakhov, "Resonant Addressing and Manipulation of Silicon Vacancy Qubits in Silicon Carbide", *Phys. Rev. Lett.* **109**, 226402 (2012).

[6] A.M. Tyryshkin, S. Tojo, J.J.L. Morton, H. Riemann, N.V. Abrosimov, P. Becker, H-J. Pohl, T. Schenkel, M.L.W. Thewalt, K.M. Itoh, S.A. Lyon, "Electron spin coherence exceeding seconds in high-purity silicon", *Nature Mat.* **11**, 143 (2012).

[7] G.W. Morley, M. Warner, A.M. Stoneham, P.T. Greenland, J. van Tol, C.W.M. Kay, G. Aeppli, "The initialization and manipulation of quantum information stored in silicon by bismuth dopants", *Nature Mat.* **9**, 725 (2010).

[8] R. Kolesov, K. Xia, R. Reuter, R. Stöhr, A. Zappe, J. Meijer, P.R. Hemmer, J. Wrachtrup, "Optical detection of a single rare-earth ion in a crystal", *Nature Comm.* **3**, 1029 (2012).

[9] J.R. Weber, W.F. Koehl, J.B. Varley, A. Janotti, B.B. Buckley, C.G. Van de Walle, D.D. Awschalom, "Quantum computing with defects", *Proc. Natl. Acad. Sciences USA* **107**, 8513 (2010).

[10] *Optical Orientation*, edited by F. Meyer and B. P. Zakharchenya (North-Holland, Amsterdam, 1984)

[11] S.E. Hayes, S. Mui, K. Ramaswamy, "Optically pumped nuclear magnetic resonance of semiconductors", *J. Chem. Phys.* **128**, 052203 (2008).

[12] I.A. Merkulov, Al.L. Efros, M. Rosen, "Electron spin relaxation by nuclei in semiconductor quantum dots", *Phys. Rev. B* **65**, 205309 (2002).

[13] D. Klauser, W. A. Coish, and Daniel Loss, "Nuclear spin state narrowing via gate-controlled Rabi oscillations in a double quantum dot", *Phys. Rev. B* **73**, 205302 (2006).

[14] S. Foletti, H. Bluhm, D. Mahalu, V. Umansky, A. Yacoby, "Universal quantum control of two-electron spin quantum bits using dynamic nuclear polarization", *Nature Phys.* **5**, 903 908 (2009).

[15] C. Kloeffel, P.A. Dalgarno, B. Urbaszek, B.D. Gerardot, D. Brunner, P.M. Petroff, D. Loss, R.J. Warburton, "Controlling the interaction of electron and nuclear spins in a tunnel-coupled quantum dot", *Phys. Rev. Lett.* **106**, 046802 (2011).

[16] E.A. Chekhovich, K.V. Kavokin, J. Puebla, A.B. Krysa, M. Hopkinson, A.D. Andreev, A. M. Sanchez, R. Beanland, M.S. Skolnick, A.I. Tartakovskii, "Structural analysis of strained quantum dots using nuclear magnetic resonance", *Nature Nanotech.* **7**, 646 (2012).

[17] A.K. Paravastu, S.E. Hayes, B.E. Schwickert, L.N. Dinh, M. Balooch, J.A. Reimer, "Optical polarization of nuclear spins in GaAs", *Phys. Rev. B* **69**, 075203 (2004).

[18] A.K. Paravastu, J.A. Reimer, "Nuclear spin temperature and magnetization transport in laser-enhanced NMR of bulk GaAs", *Phys. Rev. B* **71**, 045215 (2005).

[19] P.J. Coles, J.A. Reimer, "Penetration depth model for optical alignment of nuclear spins in GaAs", *Phys. Rev. B* **76**, 174440 (2007).

[20] Y. Li, J.P. King, L. Peng. M.C. Tamargo, J.A. Reimer, C.A. Meriles, "Helicity-independent OPNMR in Gallium Arsenide", *Appl. Phys. Lett.* **98**, 112101 (2011).

[21] J.P. King, Y. Li, C.A. Meriles, J.A. Reimer, "Optically re-writable patterns of nuclear magnetization in Gallium Arsenide", *Nature Commun.* **3**, 918 (2012).

[22] J.A. Reimer, "Nuclear hyperpolarization in solids and the prospects for nuclear spintronics", *Sol. Stat. Mag. Reson* **37**, 3 (2010).

[23] D. Paget, T. Amand, J.P. Korb, "Light-induced nuclear quadrupolar relaxation in semiconductors", *Phys. Rev. B* **77,** 245201 (2008).

[24] D. Paget, G. Lampel, B. Sapoval, "Low field electron-nuclear spin coupling in gallium arsenide under optical pumping conditions", *Phys. Rev. B* **15**, 5780 (1977).

[25] D. Paget, "Optical detection of NMR in high-purity GaAs: Direct study of the relaxation of nuclei close to shallow donors", *Phys. Rev. B* **25**, 4444 (1982).

[26] A. Malinowski, M.A. Brand, R.T. Harley, "Nuclear effects in ultrafast quantum-well spin-dynamics", *Physica E* **10**, 13 (2001).

[27] D. Mao, P.C. Taylor, "Nuclear-spin echoes in GaAs:Zn and GaAs:In", *Phys. Rev. B* **52**, 5665 (1995).

[28] O. Oda, H. Yamamoto, M. Seiwa, G. Kano, T. Inoue, M. Mori, H. Shimakura, M. Oyake, "Defects in and device properties of semiinsulating GaAs", *Semicond. Sci. Technol*. **7**, A215 (1992).

[29] J.A. Sidles, J.L. Garbini, K.J. Bruland, D. Rugar, O. Zuger, S. Hoen, C.S. Yannoni, "Magnetic resonance force microscopy", *Rev. Mod. Phys.* **67**, 249 (1995).

[30] D.C. Look, "Defects Relevant for Compensation in Semi-Insulating GaAs", *Semiconductors and Semimetals: Imperfections in III/V Materials* **38**, 91 (1993).

[31] K.L. Sauer, C.A. Klug, J.B. Miller, J.P. Yesinowski, "Optically pumped InP: Nuclear polarization from NMR frequency shifts", *Phys. Rev. B* **84**, 085202 (2011).

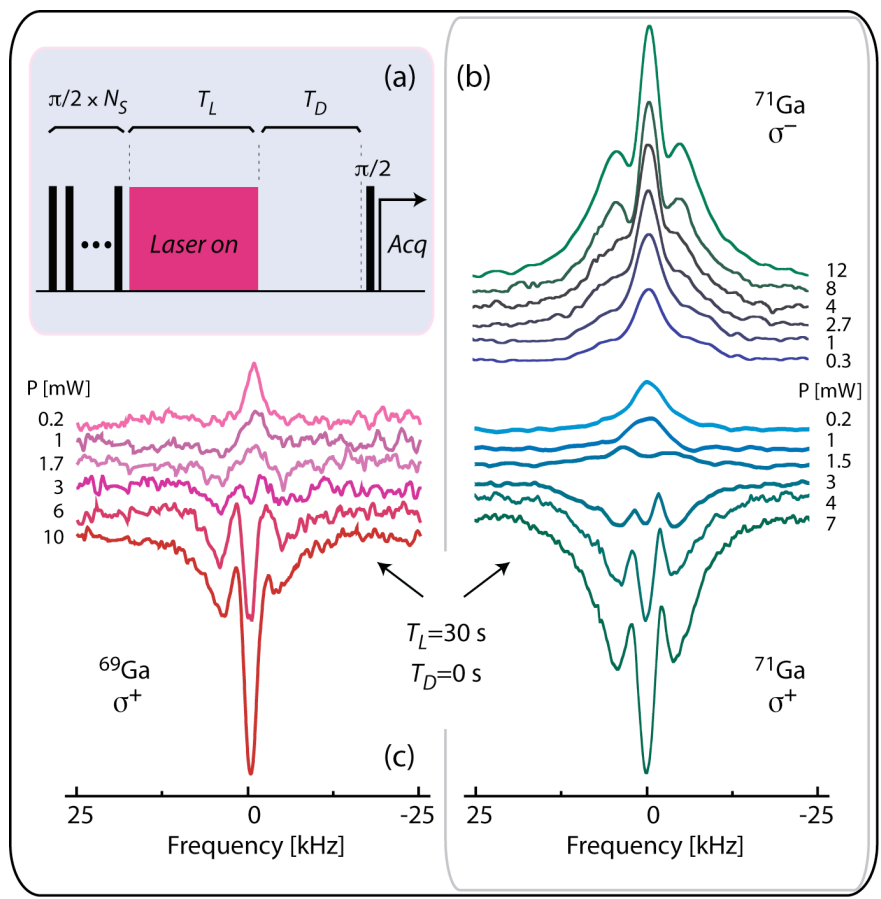

**Fig. 1:** (a) Experimental protocol. After saturating the sample magnetization with a train of $N_S$=20 radio-frequency pulses the sample is illuminated for a time $T_L$. Following a dark interval of duration $T_D$, nuclear spins are probed using a standard excitation-detection protocol. (b) $^{71}$Ga NMR signal for different laser powers; the photon helicity is $\sigma^-$ (top set) or $\sigma^+$ (bottom set). (c) $^{69}$Ga NMR response under $\sigma^+$ illumination of varying intensity. In (b) and (c) the laser wavelength is 822 nm, and the illuminated area is approximately 1 mm$^2$. The illumination time is $T_L$=30 s and signal acquisition follows immediately after, i.e., $T_D$=0 s; the temperature is 6.5 K. As the illumination intensity decreases, we observe a progressive line shape transformation marked by the disappearance of discernible satellites; in the case of $\sigma^+$ illumination we additionally witness a sign reversal of the OPNMR signal.

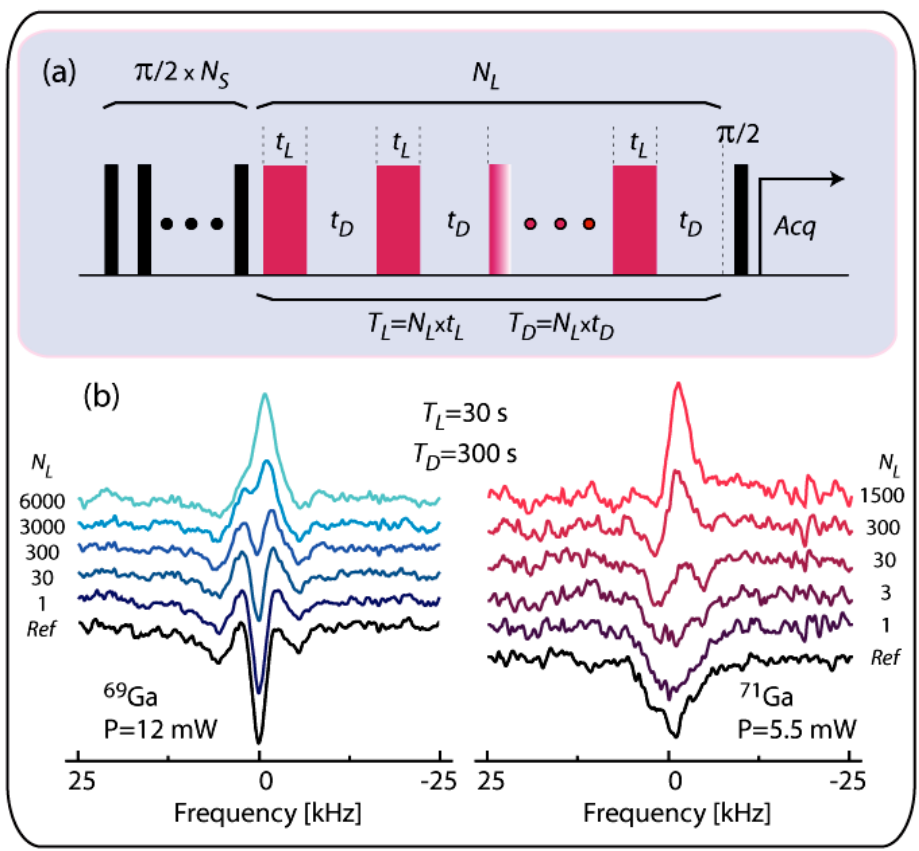

**Fig. 2:** (a) Detection protocol. While pre-saturation and detection of the magnetization remain unchanged, here we partition the preparation time $T=T_L+T_D$ into $N_L$ cycles during which light is on for a time $t_L$ and off for a time $t_D$. The total number of light pulses $N_L$ (indicated beside each trace) defines the light-on/light-off intervals within each cycle (respectively, $t_L$ and $t_D$) according to the formulas $T_L=N_L \times t_L$ and $T_D=N_L \times t_D$. (b) $^{69}$Ga (left) and $^{71}$Ga (right) NMR spectra after fixed total illumination and dark time (respectively, $T_L=30$ s and $T_D=300$ s) for a variable number $N_L$ of light-on/light-off cycles. The black traces at the bottom (labeled as 'Ref') are reference signals obtained immediately after a single 30 s long light pulse (i.e., $T_D=0$). In all cases, the laser power is 12 mW for $^{69}$Ga and 5.5 mW for $^{71}$Ga. All other conditions as in Fig. 1.

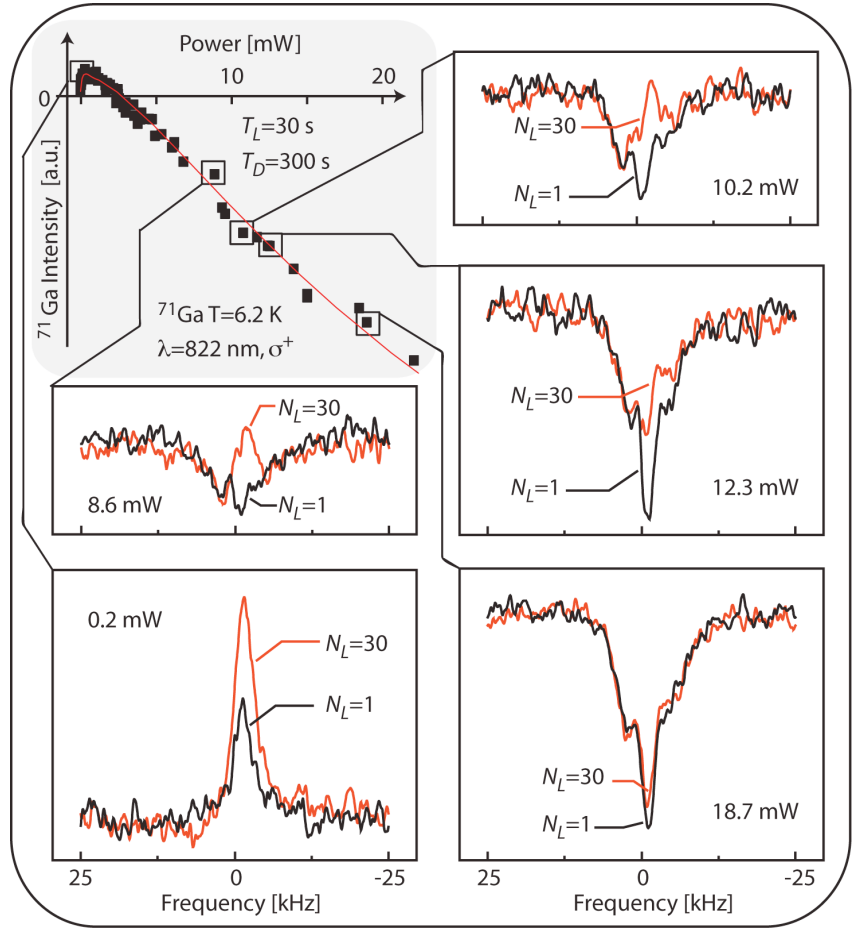

**Fig. 3:** (a) $^{71}$Ga NMR signal amplitude (central peak) as a function of laser power for the case $T_L$=30 s, $T_D$=300 s, and $N_L$=1. The red solid line is a guide to the eye. (b through f) $^{71}$Ga NMR spectrum at select laser powers. The red (black) traces correspond to the case $N_L$=30 ($N_L$=1). In all cases, $T_L$=30 s and $T_D$=300 s; all other conditions as in Fig. 1.

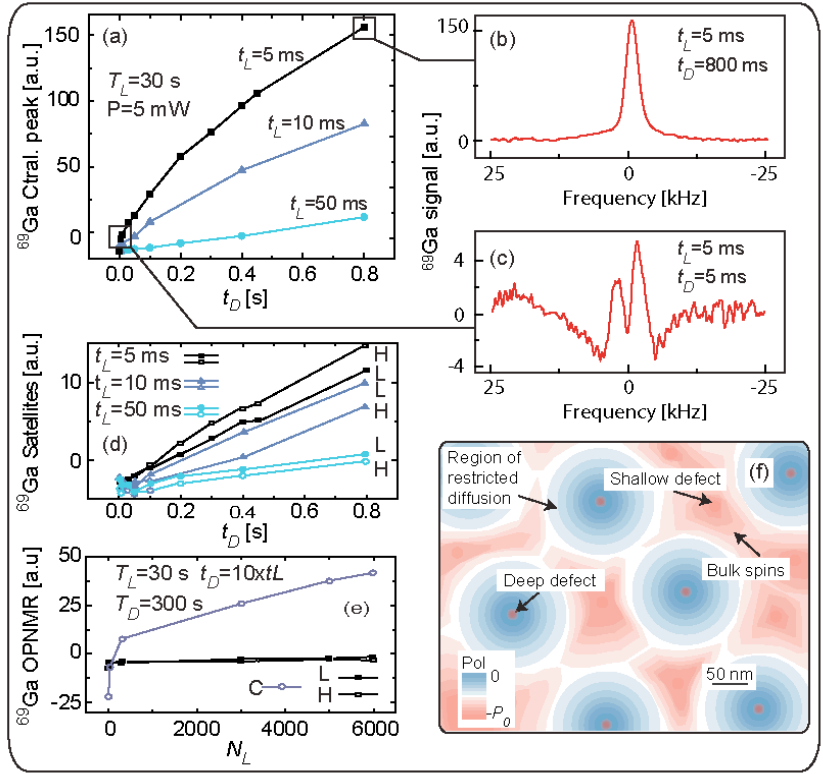

**Fig. 4:** Amplitude of the central peak in the $^{69}$Ga NMR spectrum as a function of $t_D$ for three different values of $t_L$. The total illumination time is $T_L$=30 s, the laser power is 5 mW, and the number of light-on/light-off cycles can be calculated as $N_L=T_L/t_L$ (respectively 6x10$^3$, 3x10$^3$, and 6x10$^2$ for $t_L$ equal to 5 ms, 10 ms and 50 ms). For each point the total experiment time is given by $T=T_L+T_D=N_L(t_L+t_D)$. (b,c) $^{69}$Ga NMR spectra for the conditions in (a); $t_L$ is 5 ms and $t_D$ is set to 800 ms in (b) and 5 ms in (c). (d) Satellite amplitudes in the $^{69}$Ga NMR spectrum for the conditions in (a); 'L' and 'H' respectively indicate the low and high frequency satellites. (e) Amplitude of the $^{69}$Ga central resonance peak ('C') and satellites ('L', 'H') for the conditions of Fig. 2b. (f) Representation of the distribution of nuclear polarization in the limit of high illumination intensity for σ$^+$ helicity. Nuclear magnetization (ranging from zero to some minimum, negative value –$P_0$) diffuses from shallow defects into bulk spins while nuclei near deep defects remain virtually unpolarized (blue areas). The red dots at the center of the blue circles represent the small fraction of nuclear spins near deep defects directly polarized via Fermi contact. The scale bar is approximate and is aimed only as a guide.